\documentclass{elsart}

\usepackage{graphicx}

\begin{document}
\begin{frontmatter}
\title {Design and performance of LED calibration system prototype
for the lead tungstate crystal calorimeter}
\author[IHEP]{V.A.~Batarin},
\author[FNAL]{J.~Butler},
\author[IHEP]{A.M.~Davidenko},
\author[IHEP]{A.A.~Derevschikov},
\author[IHEP]{Y.M.~Goncharenko},
\author[IHEP]{V.N.~Grishin},
\author[IHEP]{V.A.~Kachanov},
\author[IHEP]{V.Y.~Khodyrev},
\author[IHEP]{A.S.~Konstantinov},
\author[IHEP]{V.A.~Kormilitsin},
\author[IHEP]{V.I.~Kravtsov},
\author[UMN]{Y.~Kubota},
\author[IHEP]{V.S.~Lukanin},
\author[IHEP]{Y.A.~Matulenko},
\author[IHEP]{Y.M.~Melnick},
\author[IHEP]{A.P.~Meschanin},
\author[IHEP]{N.E.~Mikhalin},
\author[IHEP]{N.G.~Minaev},
\author[IHEP]{V.V.~Mochalov},
\author[IHEP]{D.A.~Morozov},
\author[IHEP]{L.V.~Nogach},
\author[IHEP]{A.V.~Ryazantsev\thanksref{addr}},
\thanks[addr]{corresponding author, email: ryazants@ihep.ru}
\author[IHEP]{P.A.~Semenov},
\author[IHEP]{V.K.~Semenov},
\author[IHEP]{K.E.~Shestermanov\thanksref{deceased}},
\thanks[deceased]{deceased}
\author[IHEP]{L.F.~Soloviev},
\author[SYR]{S.~Stone},
\author[IHEP]{A.V.~Uzunian},
\author[IHEP]{A.N.~Vasiliev},
\author[IHEP]{A.E.~Yakutin},
\author[FNAL]{J.~Yarba}
\date{\today}
\address[IHEP]{Institute for High Energy Physics, Protvino, 142281, Russian
Federation}
\address[FNAL]{Fermilab, Batavia, IL 60510, U.S.A.}
\address[UMN]{University of Minnesota, Minneapolis, MN 55455, U.S.A.}
\address[SYR]{Syracuse University, Syracuse, NY 13244-1130, U.S.A.}

\begin{abstract}

A highly stable monitoring system based on blue and red light
emitting diodes coupled to a distribution network comprised of
optical fibers has been developed for an electromagnetic calorimeter
that uses lead tungstate crystals readout with photomultiplier
tubes. We report of the system prototype design and on the results
of laboratory tests. Stability better than 0.1\% (r.m.s.) has been
achieved during one week of prototype operation.
\end{abstract}

\begin{keyword}
Light emitting diode; Monitoring system; Lead tungstate; Calorimeter; Energy
calibration\\
\PACS 29.90.+r; 85.60.Dw; 85.60.Jb; 85.60.Ha
\end{keyword}

\end{frontmatter}

\section{Introduction}

Lead tungstate (PbWO$_4$, PWO) scintillating crystals are known as
an appropriate material for use in a total absorption shower
detectors. Electromagnetic calorimeters (EMCAL) made of these
crystals have superb energy and spatial resolutions due to the
unique combination of the PbWO$_4$ physical properties~\cite{nim2}.
Several high energy physics experiments, such as ALICE and CMS at
the CERN LHC or PANDA at GSI, have decided to build their
calorimeters with the use of PWO~\cite{alice,cms,panda}. The BTeV
project at the FNAL Tevatron Collider, recently terminated by the U.
S. Dept. of Energy, intended to use these crystals~\cite{prop}.

Unfortunately, lead tungstate crystals although relatively radiation
tolerant, do lower their light output when exposed to radiation and
recover when the radiation source is removed. Extensive studies
performed at the Institute for High Energy Physics (IHEP) in
Protvino, Russia, confirmed that the PWO light output changes with
the irradiation dose rate. Dedicated measurements showed that degradation 
of light output in PWO crystals under pion irradiation with dose rates up to 
20~rad/h occurs due to light transmission loss only, rather than
changes in the scintillation mechanism~\cite{nim6}.
Further complications arise because at
the same irradiation intensity, changes in light output may vary
from one crystal to another~\cite{nim3,nim5,nim7}. In order to
maintain the intrinsic energy resolution, therefore, the system
must be continuously calibrated. In this paper, we discuss the
preferred solution for BTeV.  This technique can be applied for any
detector with similar operational conditions.

The BTeV calorimeter was designed to cover the space with a radius
1.6~m near the beam axis, about 220 mr of angle from the interaction
point. There were approximately 10,000 PWO crystals coupled with
photomultiplier tubes (PMT). About 90\% of crystals would suffer from 
radiation with dose rates less than 20 rad/h. 
The expected energy resolution of the EMCAL was $1.7\%/\sqrt{E}\oplus0.55\%$, 
and the accuracy of the energy calibration should be better than 0.2\%.
Monte-Carlo studies show that electrons and positrons produced in
physics events, mainly from semileptonic B-decays or from photon
conversions near the interaction region, can be successfully used to
calibrate the detector in-situ~\cite{tdr}. The amount of time required to
collect sufficient samples would be significantly vary in different
areas of the EMCAL but even in the worst case scenario would not
exceed one day period. However, the calorimeter would need to be
continuously monitored within these time intervals or during
Tevatron shutdown periods.  


In addition to the crystals light output change, PMT gain
instabilities could deteriorate the performance. A usual way to
track the PMT gain variations is the use of a monitoring system with
a light pulser. If a light pulse could be sent to the PMTs
directly, it would be relatively easy to measure PMT gain changes.
However, in our case the crystals more than cover the entire
detection surface of the mating PMTs; thus the only solution would
be to send
light to the PMT photocathodes through the crystals. Therefore the
same monitoring system that is used to measure the radiation effects
needs to be used to monitor the PMTs.

%
To monitor crystal light output changes, we use a blue light pulser
with a wavelength close to the 430~nm emission peak of the PbWO$_4$
crystal. Since these light pulses are detected by PMTs, what we
measure is the change in the product of the PMT gain and the crystal
transparency. To monitor the PMT gain changes we use a red light
pulser, since the red light transmission in the crystals changes
much less due to radiation than the blue light
transmission~\cite{red}. In our test beam studies, the separation of
these two sources of signal variations was crucial and allowed us to
study the changes in the crystal properties alone. Our experience
with a blue-red light pulser system at the test beam facility is
discussed in~\cite{nim3}.

Taking into account the conditions described above we can summarize
the main requirements for the monitoring system light pulser:\\
- high luminous flux for red and blue (close to 430~nm) light pulses
to be able illuminate at least 2600 fibers of the light distribution
network
providing PMT signals equal to those from 20 GeV electrons;\\
- non-uniformity of the light spot illuminating the bunch of fibers
should be not more than 10\%;\\
- stability at the level of $2\cdot$10$^{-3}$ over a day.

We decided to design a monitoring system with the use of light
emitting diodes since LED pulsers provide a very reliable operation
and required stability as it was shown in~\cite{nim4}. The whole
system should consist of four identical LED pulser modules, each
monitoring a quarter of calorimeter. Only one module would be
powered in a given time interval. This solution allowed to stay
within the bandwidth of the data acquisition system (DAQ) while
collecting monitoring data. The prototype of such module was
designed and tested at the Institute for High Energy Physics in
Protvino.

\section{Prototype Design}

\begin{figure}
\centering
\includegraphics[width=0.95\textwidth]{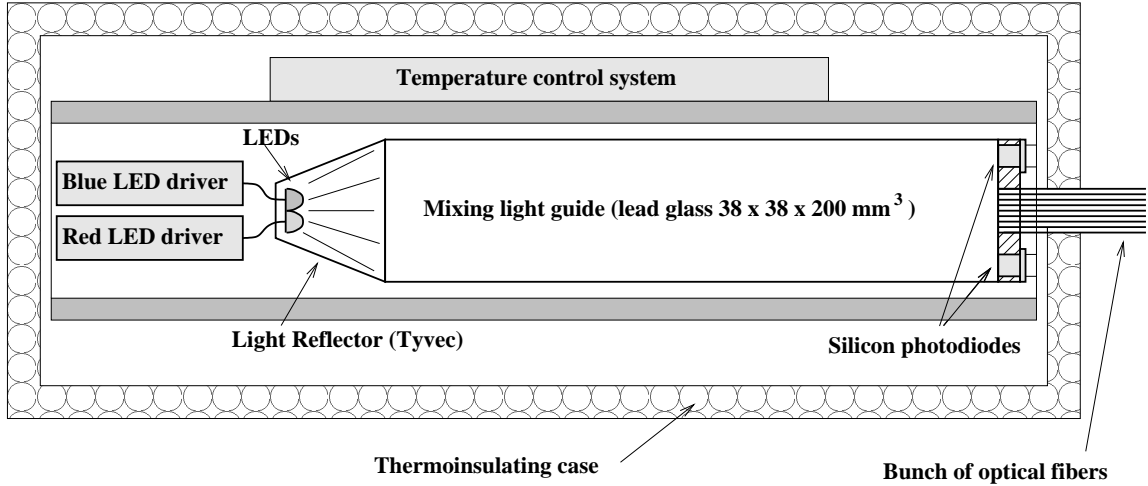}
\caption{Block diagram of the LED pulser.}
\label{fig:pulser}
\end{figure}

The light pulser prototype is shown schematically in Fig.~\ref{fig:pulser}.
The system includes:
\begin{itemize}
\item blue and red LEDs;
\item two LED drivers;
\item light reflector;
\item mixing light guide;
\item two reference silicon photodiodes;
\item bunch of optical fibers;
\item temperature control system;
\item thermoinsulating case.
\end{itemize}

Powerful blue and red LEDs from Lumileds Lighting, USA, illuminate
optical fibers and reference photodiodes through a bar of lead glass
with the dimensions of $38\times 38\times 200$~mm$^2$ which was used
as a light mixer. To improve light collection in the mixer, LEDs
were placed inside a square pyramid with a reflecting internal
surface near the apex. The cross-section of the light mixer allows
to illuminate simultaneously about 3000 optical fibers of 0.4 mm
diameter. We decided to use silica fibers FIL300330370 by Polymicro
Technologies, USA~\cite{fiber}. They have a core of 300 micron
diameter and an aluminium buffer providing excellent mechanical
strength. According to the results of the radiation hardness
measurements with a $\gamma$-source obtained by the CMS ECAL group,
these fibers keep their light transmittance at the constant level up
to 12 Mrad of absorbed dose~\cite{fibers_cms}. This is very
important because some part of fibers would be irradiated with high
dose rates during the setup operation.

Technical parameters of the LEDs are given in
Table~\ref{tab:leds}~\cite{luxeon}.
Besides the exceptional luminous fluxes, we found that two additional
features of the Luxeon technology are very important for our purposes:
very long operating life (up to 100,000 hours in DC mode) and small
temperature dependence of the light output (less than 0.5\%/$^\circ$C).

\begin{table}[h]
\begin{center}
\caption{The properties of LEDs used in the light pulser~\cite{luxeon}.}
\begin{tabular}{|l|c|c|}
\hline
LED Property                   & LXHL-PR02        & LXHL-PD01      \\
                               &(Royal Blue Radiometric) & (Red)   \\
\hline
Brand                          & Luxeon V Emitter & Luxeon Emitter \\
Typical Luminous Flux          &                  & 44 lm (@350 mA)\\
Typical Radiometric Power      & 700 mW (@700 mA) &                \\
Radiation Pattern              & Lambertian       & Lambertian     \\
Viewing Angle                  & 150 degrees      & 140 degrees    \\
Size of Light Emission Surface & $5\times 5$ mm$^2$ & $1.5\times 1.5$ mm$^2$\\
Peak Wavelength                & 455 nm           & 627 nm         \\
Spectral Half-width            & 20 nm            & 20 nm          \\
Maximum DC Forward Current     & 700 mA           & 350 mA         \\
\hline
\end{tabular}
\label{tab:leds}
\end{center}
\end{table}

\begin{figure}
\centering
\includegraphics[width=0.95\textwidth]{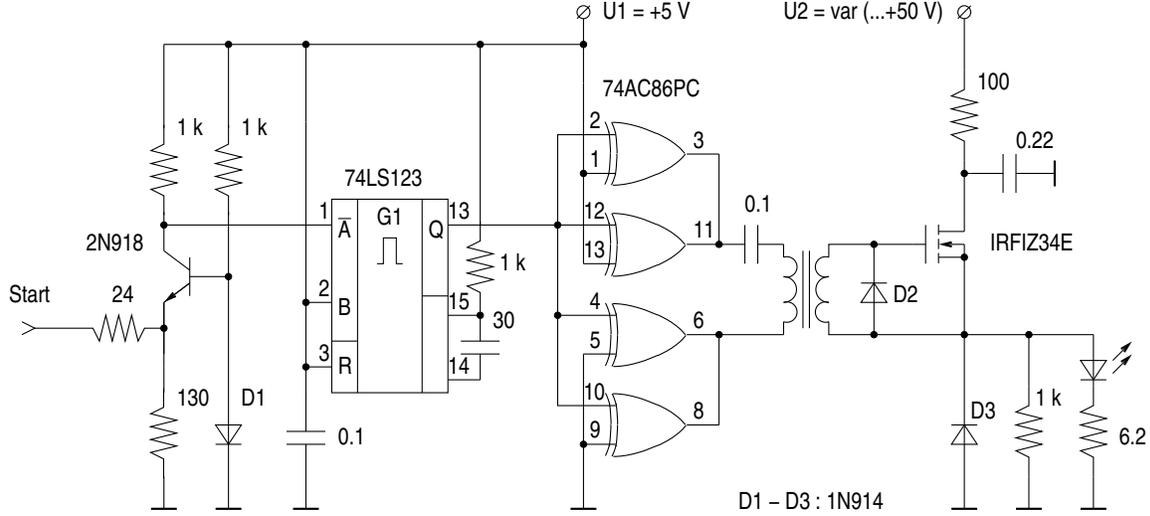}
\caption{LED driver circuit.}
\label{fig:circuit}
\end{figure}

The electronic circuit for the LED driver is shown in
Fig.~\ref{fig:circuit}. The drivers of red and blue LEDs are
identical. They are triggered by pulses of standard NIM-logic
levels. Each driver includes a shaping amplifier determining the
duration of the light flashes and an output powerful transistor (MOS
FET). The transistor switches LED to a voltage source adjustable in
a range up to +50~V which allowed us to tune the necessary
brightness of the light pulses.

An  essential element of the light monitoring system is a stable
reference photodetector with a good sensitivity at short wavelengths
which measures light pulses amplitude variation in time.
Silicon PN-photodiodes S1226-5BQ by Hamamatsu, Japan,
are well suited to this task because they have high ultraviolet and
suppressed infrared sensitivity, low dark current and small temperature
coefficient (less then 0.1\%/$^\circ$C) in the range of wavelengths from
200 to 700~nm~\cite{ham}.
The rather large (about 6~mm$^2$) sensitive area of this photodiode allows
us to work without preamplifiers, thus, improve a stability of the reference
system itself. In the prototype, we used two photodiodes attached to the
output window of the light mixer in the corners.

Our previous studies showed that temperature variations deteriorate the
performance stability of the LED monitoring system~\cite{nim4}.
Therefore we designed a heat insulated case with a possibility
to control temperature inside it. A simple electronic circuit with a
thermistor in the feedback has been placed in the same case. The operating
temperature inside the case should be higher than expected maximum of the
room temperature since the system contains only heaters. We expected that
temperature variation in the BTeV experimental hall would be relatively
small (few degrees) over the data taking period, so the suggested solution
is adequate.

\section{Test setup}

Our test setup consisted of the LED pulser prototype and a lead
tungstate crystal coupled with a PMT Hamamatsu R5800, all placed in
a light-tight box. Instead of a fiber bunch, we used one silica
optical fiber to transport light from the output window of the light
mixer to the crystal edge. The crystal and the PMT were taken from
the $5\times 5$ calorimeter prototype tested with a beam earlier.
Thus, we knew approximate correspondence between an energy of
electrons hitting the crystal and amplitude of the anode signal for
the fixed PMT gain.
The DC voltage source was common for two LED pulsers. Its output
level was set to the value which gave the PMT anode signal from the
blue LED equivalent to that from a 20 GeV electron.

Our data acquisition system was described in detail in~\cite{nim2}.
We used LeCroy~2285 15-bit integrating ADC
to measure signal charges from the PMT and photodiodes over 150 ns gate.
Besides, temperature was measured continuously during data taking with the
use of five thermosensors placed in different locations. One of them
provided an information about room temperature, another one was
installed near the photocathode window of the PMT. Three other
sensors performed temperature measurements inside the prototype case,
namely: near the LEDs, near the photodiodes and at the surface of the heater.

\section{Experimental results}
\subsection{Light spot uniformity}

\begin{figure}
\centering
\includegraphics[width=0.6\textwidth]{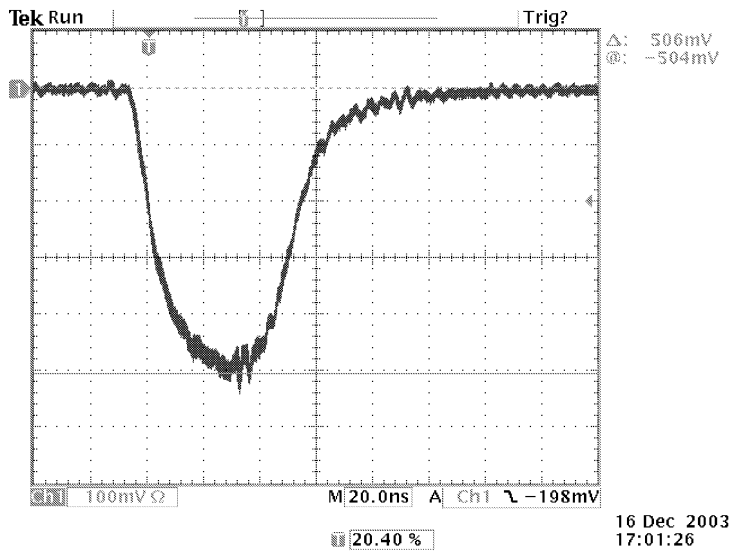}
\caption{Blue LED signal at the anode of the PMT.}
\label{fig:oscill}
\end{figure}

\begin{figure}
\centering
\includegraphics[width=0.6\textwidth]{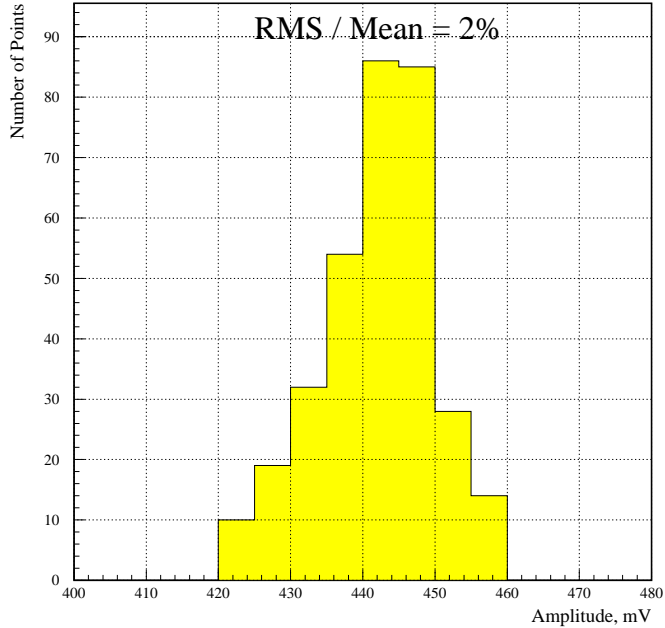}
\caption{Uniformity of the light distribution over the output window of the
light mixer.}
\label{fig:uniformity}
\end{figure}

Uniformity of the light distribution over the output window of the light
mixer was measured by means of manual surface scan accomplished with a single
optical fiber with the step size of 2~mm. The scan area was
$34\times 34$~mm$^2$. Light signal was detected by the PMT and the pulse
heights were measured with a scope. The shape of the blue LED signal at the
anode of the PMT is shown in Fig.~\ref{fig:oscill}.
The distribution of measured pulse heights is shown in
Fig.~\ref{fig:uniformity}. The r.m.s. of this distribution is 2\%,
and the full width is 9\%.

\subsection{Temperature dependence}

In order to estimate the temperature
dependence of the light pulser prototype components, we performed
measurements with two different temperatures inside the case, 27$^\circ$C and
45$^\circ$C. During these measurements the temperature of
 the PMT remained stable, and we compared the signals measured by this PMT.
The mean ADC count of the blue LED signal distribution became
smaller by 11.5\% when the temperature increased from 27$^\circ$C to
45$^\circ$C.
Assuming that the temperature dependence is linear, the coefficient
is estimated as -0.64\% at 27$^\circ$C. The same analysis was done
for the same LED signals measured by the photodiodes. We averaged
between the results of measurements performed by each photodiode and
obtained the temperature coefficient of the system blue LED pulser -
photodiode is equal to -0.60\%/$^\circ$C. This means that photodiodes
have their own temperature coefficient about 0.04\%/$^\circ$C in the
region of 455~nm wavelength. The measured temperature coefficient of
the red LED pulser is -1.0\%/$^\circ$C, and that of the photodiodes
in the red region is 0.2\%/$^\circ$C at 27$^\circ$C. The obtained
results show that to keep a stability of the whole system at the
level better than 0.2\% we should reduce the temperature variation
near the LEDs and the photodiodes down to 0.2$^\circ$C.

\subsection{Long-term stability}

To evaluate stability of the light pulser prototype we collected
data continuously over one week. In these measurements, the DAQ
recorded the following information every 9 seconds: 10 pulse heights
from each LED detected by the photodiodes and by the PMT as well as
the temperature data. For the analysis, we calculated mean values of
signals accumulated over consequent 20 minute time intervals and
formed their distributions. The r.m.s. of such distribution
characterizes the stability of the given signal over the period of
measurements.

\begin{figure}
\centering
\includegraphics[width=0.7\textwidth]{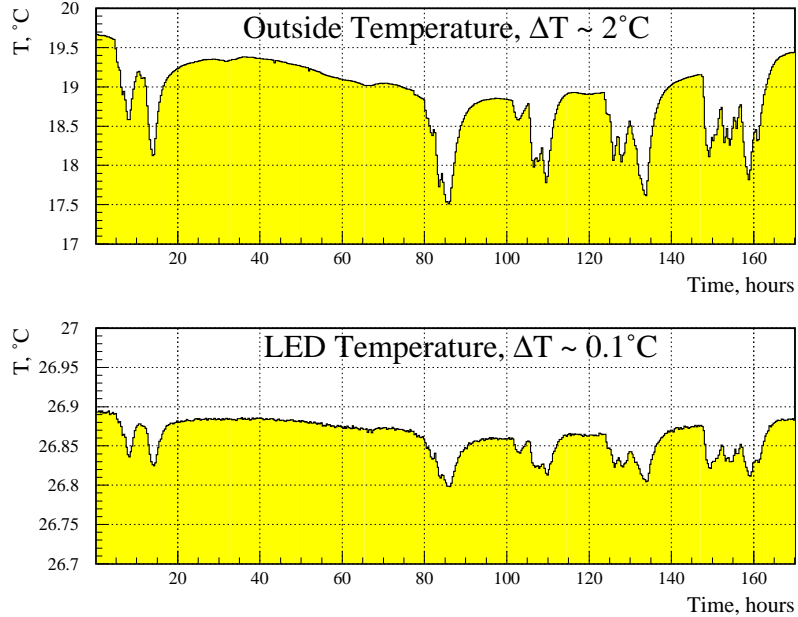}
\caption{Temperature variations during the prototype stability test.}
\label{fig:tempstab}
\end{figure}

Figure~\ref{fig:tempstab} shows a variation of room temperature and
temperature in the region of LED's over one week . We can see that
temperature of LED's was stable within 0.1$^\circ$C while the change
of temperature outside the case achieved 2$^\circ$C.

\begin{figure}
\centering
\includegraphics[width=0.7\textwidth]{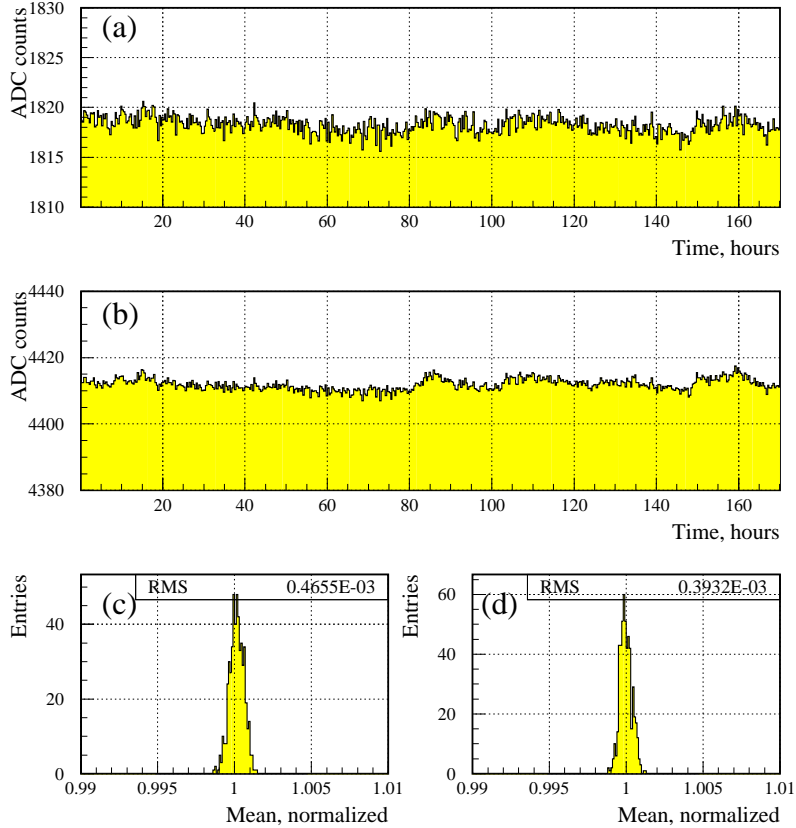}
\caption{Stability of the LED pulser prototype: (a),(b) - dependence
in time of blue and red LED signals respectively detected by one of
the photodiodes over one week of measurements; each entry is a mean
value of amplitude distribution collected over 20 min; (c),(d) -
distributions of mean values; r.m.s. characterizes stability of the
system over one week.} \label{fig:stability}
\end{figure}

The dependence in time of blue and red LED signals detected by one
of the reference photodiodes over one week of measurements is shown
in Fig.~\ref{fig:stability}(a) and (b) respectively. Normalized
distributions that allow to evaluate stability of the whole system,
i.e. the LED pulsers and the photodiode, are given in
Fig.~\ref{fig:stability}(c) for blue and \ref{fig:stability}(d) for
red LEDs. The r.m.s. of these distributions, expressed in percent,
are 0.05\% and 0.04\%. As expected, outside temperature variation
didn't affect the performance of the prototype.

\section{Summary}

We have developed the LED-based monitoring system for the
electromagnetic calorimeter that uses PWO crystals coupled with PMTs.
The expected conditions and demands of the BTeV project were taken into
account. The prototype of the light pulser based on the blue and red LEDs and
reference silicon photodiodes has been designed, assembled and succesfully
tested in the laboratory.

The prototype module is capable to provide continuous monitoring of
the PMTs gain variation and crystals light output change due to the
beam irradiation for about 3000 cells of the EMCAL. The maximum
difference of the light pulses intensity in different channels is
9\%. The prototype stability was estimated over the time period of
one week. We found that the blue LED pulser is stable to 0.05\% and
the red LED pulser is stable to 0.04\%, within one week of
continuous operation. This exceeded the requirements of the project.

This highly stable monitoring system combined with in-situ
calibration of the EMCAL would ensure the superb intrinsic
resolution of the lead tungstate crystal calorimeter over the whole
period of its operation.

\section{Acknowledgements}

We thank the IHEP management for providing us infrastructure
support. Special thanks to Fermilab for providing equipment for data
acquisition. This work was partially supported by U.S. National
Science Foundation and U. S.
Department of Energy and the Russian
Foundation for Basic Research grant 02-02-39008.


\end{document}